\documentclass{appolb}
\usepackage{epsfig}
\usepackage{lscape}
\usepackage{color}
\usepackage{here}
\usepackage{amsmath}
\usepackage{amssymb}
\usepackage{graphicx}
\usepackage{graphics}

 \def\be{\begin{equation}}
 \def\ee{\end{equation}}
 \def\bea{\begin{eqnarray}}
 \def\eea{\end{eqnarray}}
 \def\gsim{\mathrel{\rlap{\lower0.2em\hbox{$\sim$}}\raise0.2em\hbox{$>$}}}
 \def\ksim{\mathrel{\rlap{\lower0.2em\hbox{$\sim$}}\raise0.2em\hbox{$<$}}}
 \def\kg{\mathrel{\rlap{\lower0.25em\hbox{$>$}}\raise0.25em\hbox{$<$}}}

\begin{document}
\eqsec 

\title{Radiative and Collisional Energy Loss of Heavy Quarks in Deconfined Matter}
\author{ J. Aichelin\footnote{invited speaker},  P.B. Gossiaux,T. Gousset
\address{SUBATECH, Universit\'e de Nantes, EMN, IN2P3/CNRS,\\
 4 rue Alfred Kastler, 44307 Nantes cedex 3, France}}
\maketitle
\begin{abstract}
We extend our recently advanced model on collisional energy loss of heavy quarks 
in a quark gluon plasma (QGP) by including radiative energy loss. We discuss the approach 
and present calculations for PbPb collisions at $\sqrt{s}=2.76 TeV$. The transverse momentum spectra, 
RAA, and the elliptic flow $v_2$ of heavy quarks have been obtained using the model of Kolb and Heinz
for the hydrodynamical expansion of the plasma.
\end{abstract}

\section{Introduction}
There are many pieces of evidence that in ultrarelativistic heavy ion collisions a plasma of quarks and gluons (QGP) is formed. Such a state of matter, in which unbound gluons and quarks are in local thermal equilibrium, has been predicted by
lattice gauge calculation, the numerical solution of the Lagrangian of Quantum Chromodynamics (QCD) on a lattice. 
 
In heavy ion reactions such a state can be only created for a very short time (of the order of $10^{-23}s$). Then the
system, which expands with almost the speed of light, forms hadrons which are finally observed in the detector. The
problem is now to conclude from the observed hadrons on the existence and the properties of such a QGP. This is all but easy.  It turned out that the multiplicity of light hadrons is very well described by statistical models. This means that at the end of the plasma expansion, when hadrons are formed, the system is in thermal equilibrium and therefore light hadrons do not carry information on the early stage of the expansion.

Therefore only few observables can be used to study the interior of the plasma.
They include the high $p_t$ hadrons which originate from jets a as well as the $p_t$ and $v_2$ distribution of heavy mesons which contain either a c or a b quark because neither jets nor heavy quarks come to an equilibrium with the plasma.

Heavy quarks are produced in hard binary initial collisions between the incoming protons. Their production cross sections
are known from pp collisions and can as well be calculated in pQCD calculations. Therefore the initial transverse momentum distribution of the heavy quarks is known. Comparing this distribution with that measured in heavy ion collisions allow to define
$R_{AA}=(d\sigma_{AA}/dp_t^2)\ / \ (N_c\, d\sigma_{pp}/dp_t^2)$, where $N_c$ is the number of the initial binary collisions between
projectile and target. The deviation of $R_{AA}$ from one measures the interaction of the heavy quark with the plasma because the hadron cross sections of heavy mesons are small. The heavy quark does not come to thermal equilibrium with the QGP
therefore $R_{AA}$ contains the information on the interaction of the heavy quark while it traverses the plasma. In addition,
the distribution of heavy quarks at the moment of their creation is isotropic in azimuthal direction, therefore the elliptic flow
$v_2 = <cos2(\phi- \phi_R)>$, where $\phi$ ($\phi_R $) is the azimuthal angle of the emitted particle (reaction plane) is 0. The
observed  finite $v_2$ value of the observed heavy meson can only originate from interactions between light QGP constituents
and the heavy quarks. The simultaneous description of $R_{AA}$ and $v_2$ and their centrality dependence, presently the only observables for which data exist, give then the possibilities to understand the interactions inside the QGP.

Unfortunately the experimental results depend not only on the elementary interaction but also on the description of the expansion of the QGP \cite{Gossiaux:2011ea}. Therefore the ultimate aim is to control the expansion by results on the light meson sector. This has not been achieved yet for the LHC and therefore it is difficult to asses the influence of the expansion on the observables. We use here the approach from Kolb and Heinz which has reasonably well described the midrapidity light mesons at RHIC \cite{Kolb:2003dz}. We adjust only the charged particle multiplicity to the value measured at LHC.

The $R_{AA}$ of 0.2 values observed for large $p_t$ heavy mesons are much smaller than originally expected. Early theoretical
approaches based on perturbative QCD (pQCD) calculation gave much larger values and it has been doubted, whether pQCD is the right tool to describe this interaction. This early calculation, however, used ad hoc assumptions on the coupling constant $\alpha_s$ and
the infrared regulator $\mu$. With a standard choice $\mu$ and $\alpha_s$ an artificial K factor, an overall multiplication factor of the elastic cross section of around 10  \cite{Moore:2004tg,Svetitsky:1987gq} had to be introduced to
match the experimental data.

A while ago we advanced an approach for the collisional energy loss of heavy quarks in the QGP \cite{Gossiaux:2008jv,Gossiaux:2009mk,Gossiaux:2009hr}
in which a) $\mu$ has been fixed by the demand that more realistic calculations using the hard thermal loop approach give the same energy loss as our Born type pQCD calculation and b) the coupling constant is running and fixed
by the sum rule advanced by Dokshitzer and later used by Peshier. Both these improvements increased the cross section especially for small momentum transfers and reduced therefore the necessary K factor to 2. Here we include in addition
the radiative energy loss \cite{Gossiaux:2010yx,agg}.

\section{Model}
\begin{figure}
\begin{center}
\epsfig{file=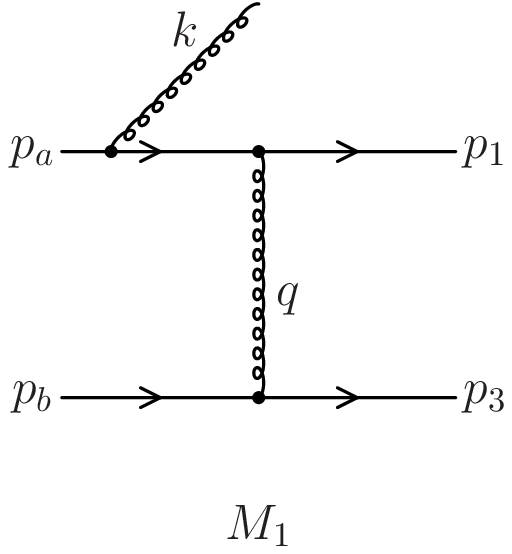,width=0.3\textwidth}
\epsfig{file=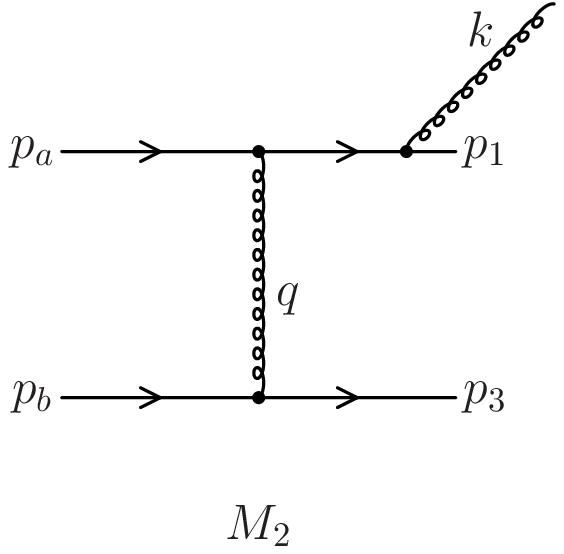,width=0.3\textwidth}
\epsfig{file=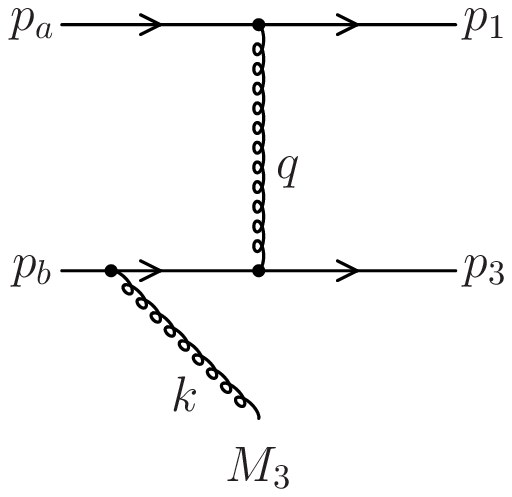,width=0.3\textwidth}
\epsfig{file=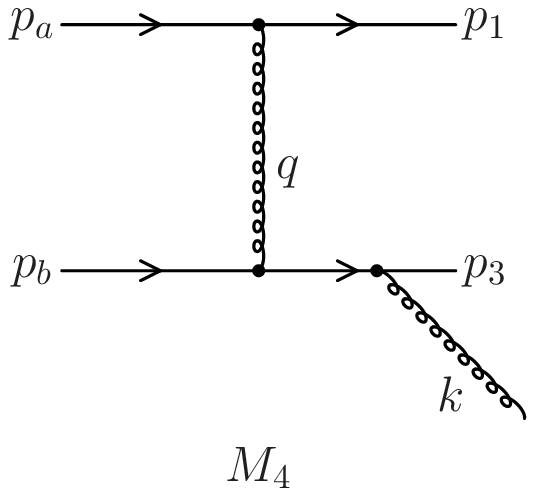,width=0.3\textwidth}
\epsfig{file=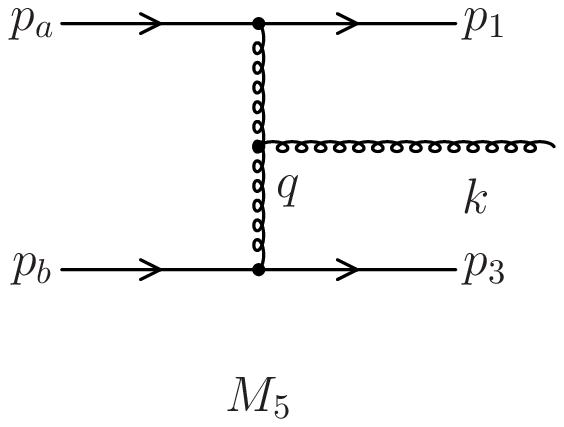,width=0.3\textwidth}
\end{center}
\caption{(Color online) The five matrix elements which contribute to the gluon bremsstrahlung.}
\label{dia}
\end{figure}

Our approach extended by including radiative energy loss has been well described the heavy quark data at RHIC. Therefore it is worthwhile to calculate what we expect for LHC energies if we modify teh model only in a minimal way by
adjusting the initial condition do $dN/dy = 1600$, as observed at RHIC. To include radiation
we have to consider the following 5 matrix elements, displayed in fig. \ref{dia}, which contributes to radiation.
The commutation relation \be T^bT^a=T^aT^b-if_{abc}T^c\label{tm}\ee allows us to regroup the 5 matrix elements into
3 combinations, each of them being independently gauge invariant:
\bea iM^{QED}_{h.q.} &=& C_a i(M_1+M_2)\nonumber \\
iM^{QED}_{l.q.} &=& C_a' i(M_3+M_4)\nonumber \\
iM^{QCD}&=& C_c i(M_1+M_3+M_5).
\eea
h.q. (l.q.) mark the emission of the gluon from the heavy (light quark) line. $C_a$, $C_a'$ and $C_c$ are the color algebra
matrix elements. The matrix elements labeled as QED are the bremsstrahlung diagrams already observed in Quantum Electrodynamics (QED), whereas that labeled QCD is the genuine diagram of Quantum Chromodynamics (QCD). The QCD diagram is the main objet of interest here because it dominates the energy loss of heavy quarks.
\begin{figure}
\begin{center}
\epsfig{file=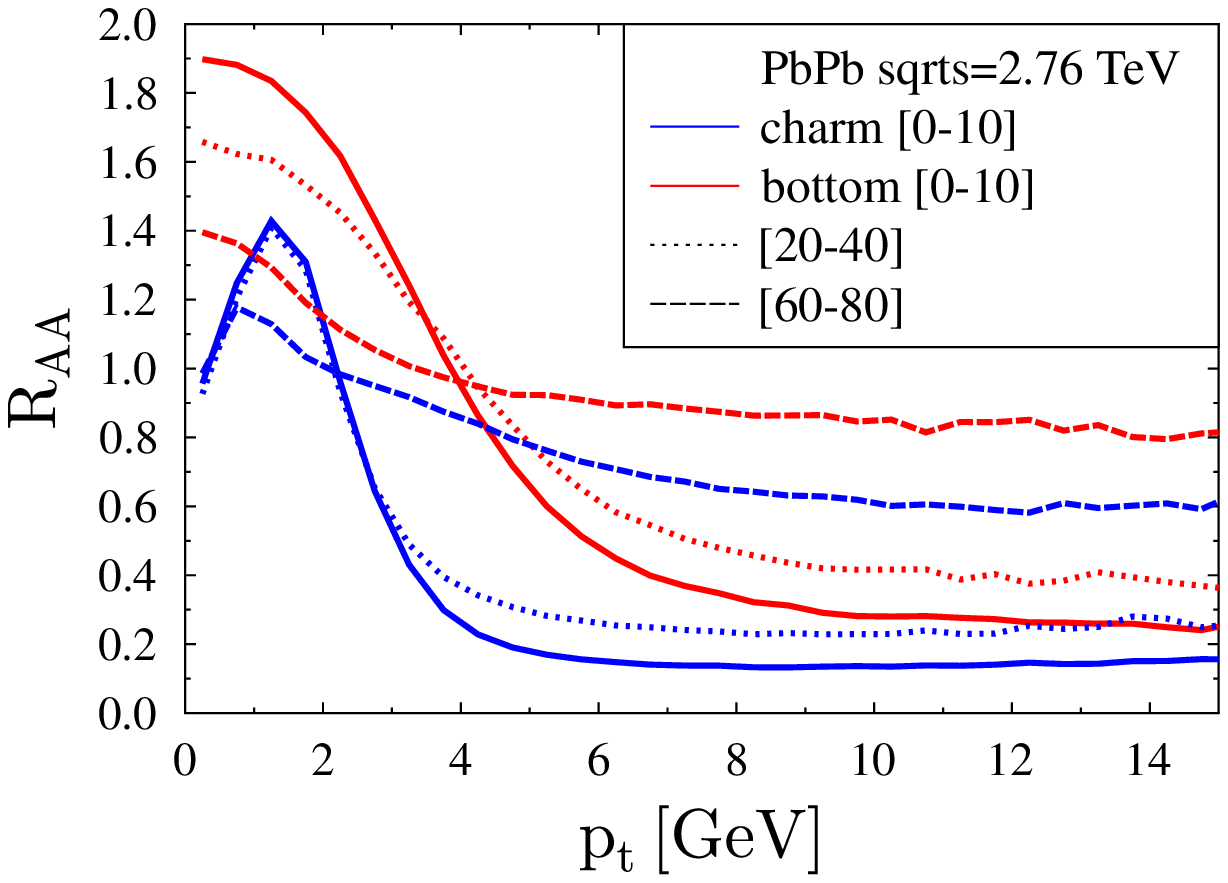,width=0.55\textwidth}
\epsfig{file=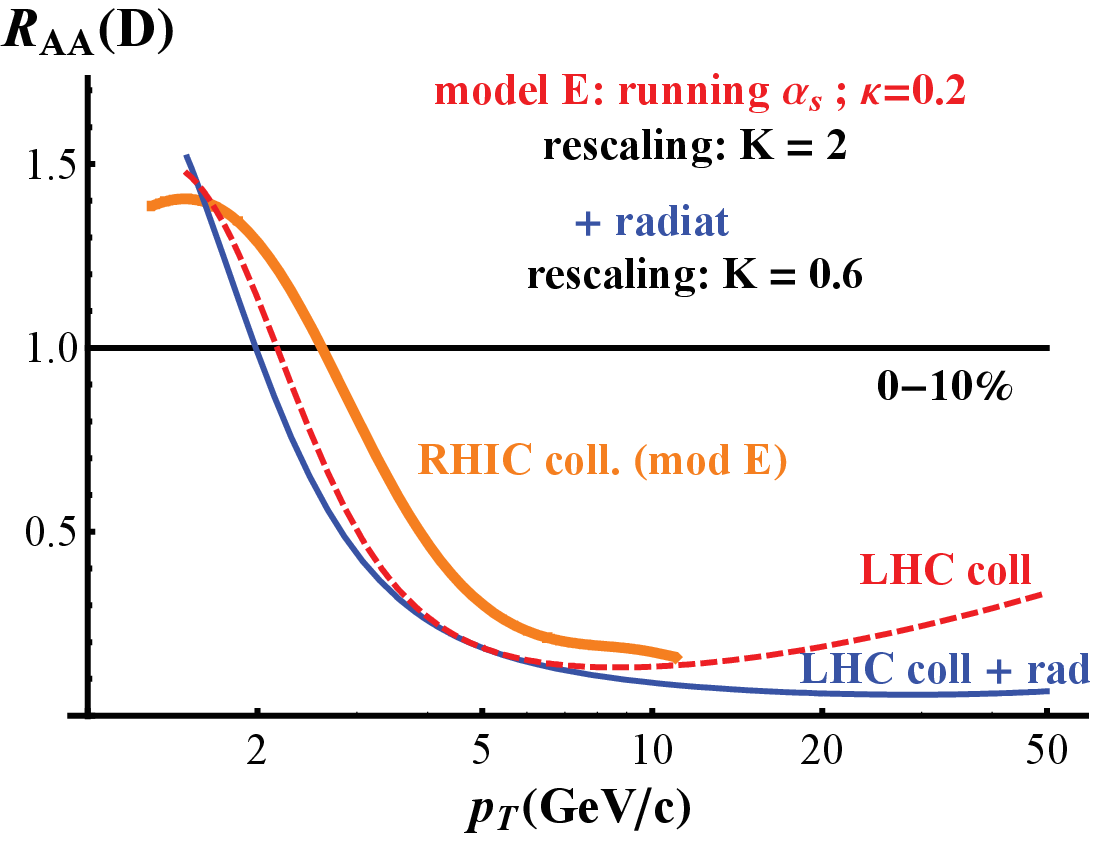,width=0.45\textwidth}
\epsfig{file=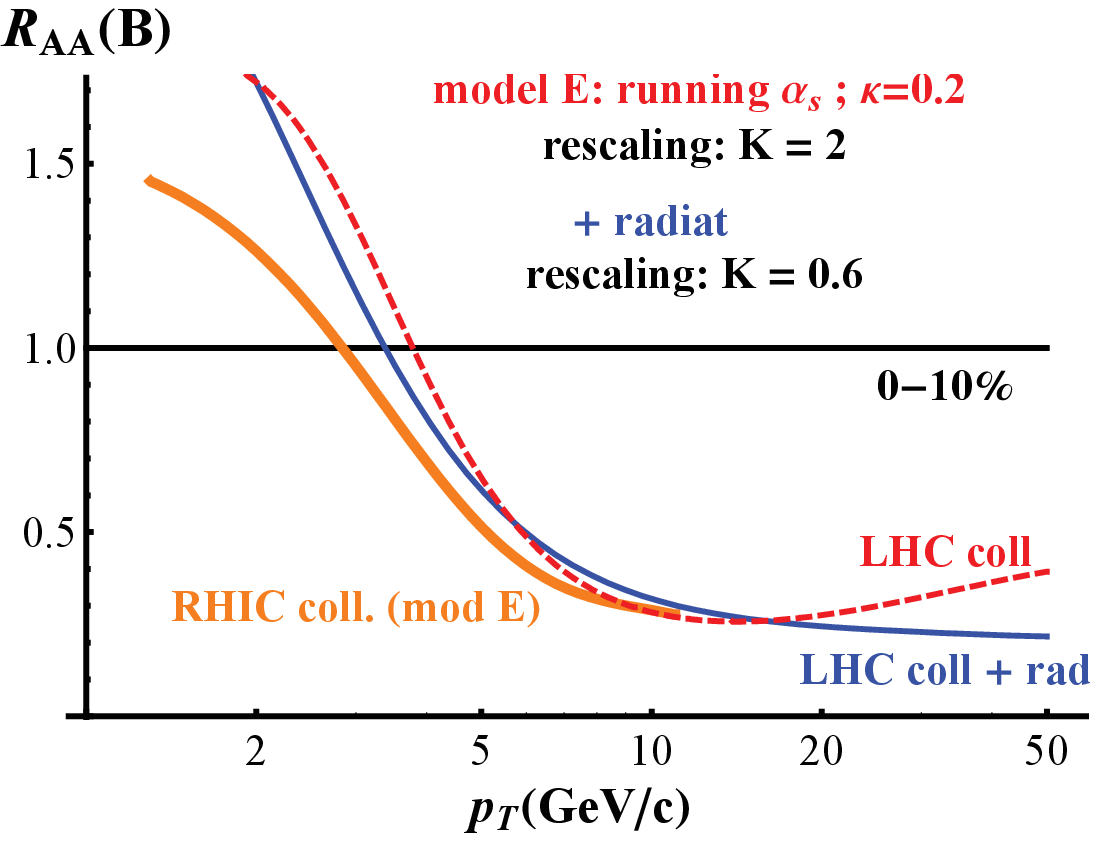,width=0.45\textwidth}
\end{center}
\caption{(Color online) The transverse momentum distribution of $R_{AA}$ at midrapidity for different centralities and 
for bottom (blue)  and charm (red) quarks. Top: The details at small $p_t$, bottom left (right):  $R_{AA}$ at large $p_t$ 
for D-mesons (B-mesons). }
\label{pt}
\end{figure}
We evaluate the matrix elements in scalar QCD (see ref.\cite{Meggiolaro:1995cu}). They are given by
\bea iM_1^{SQCD} &=& C_A(ig)^3
\frac{(p_b+p_3)^\mu}{(p_3-p_b)^2}D_{\mu\nu}[p_3-p_b]
\big( \frac{(p_a+p_1-k)^\nu
(2p_a-k)\epsilon}{(p_a-k)^2-m^2} - \epsilon^\nu \big)
\nonumber\\
iM_5^{SQCD}&=&C_c(ig)^3
D^{\mu\mu'}[p_3-p_b]D^{\nu\nu'}[p_1-p_a]\left[g_{\mu'\nu'}(p_a-p_1+p_3-p_b)_\sigma+
\right.\nonumber\\
&& \left. g_{\nu'\sigma}(p_1-p_a-k)_{\mu'}+g_{\sigma\mu'}(p_b-p_3+k)_{\nu'}\right]
\epsilon^\sigma \nonumber \\
&\cdot&\frac{(p_3+p_b)^\mu(p_a+p_1)^\nu}
{(p_3-p_b)^2(p_1-p_a)^2}
\eea
$M_3$ is obtained by replacing $p_a \to p_b$ and $p_1\to p_3$ in $M_1$. Using light cone gauge and keeping only
the leading term in $\sqrt{s}$ we find that the square of the matrix element factorizes
\be
|M|^2 = |M_{elast}(s,t)|^2 P_g(m,t,\vec{k_t},x)
\ee
with  $|M_{elast}(s,t)|^2 = g^4\frac{4s^2}{t^2}$  being the matrix element squared for the elastic
cross section in a coulomb-like interaction between the heavy quark and a light quark (gluon).
$P_g(m,t,s,\vec{k_t})$ describes the distribution function of the produced gluons. 
To discuss the physics we adopt the following light cone vectors\bea
p_a&=&\{\sqrt{s-m^2},\frac{m^2}{\sqrt{s-m^2}},0,0\}\nonumber \\
p_b&=&\{0,\sqrt{s-m^2},0,0\}\nonumber \\
k&=&\{x\sqrt{s-m^2},0,\vec{k_t} \}\nonumber \\
p_1&=&p_a+q-k=\{p_a^+(1-x)-\frac{q^2_t}{p_b^-},
\frac{(\vec{k_t}-\vec{q_t})^2+m^2}{(1-x)p_a^+},\vec{q_t}-\vec{k_t}\}\nonumber \\
p_3&=& p_b-q = \{\frac{q_t^2}{p_b^-},p_b^--
\frac{(1-x)k_t^2-x(\vec{k_t}-\vec{q_t})^2+m^2x^2}{p_a^+(1-x)x},-\vec{q_t}\}\nonumber \\
\eea
\begin{figure}
\begin{center}
\epsfig{file=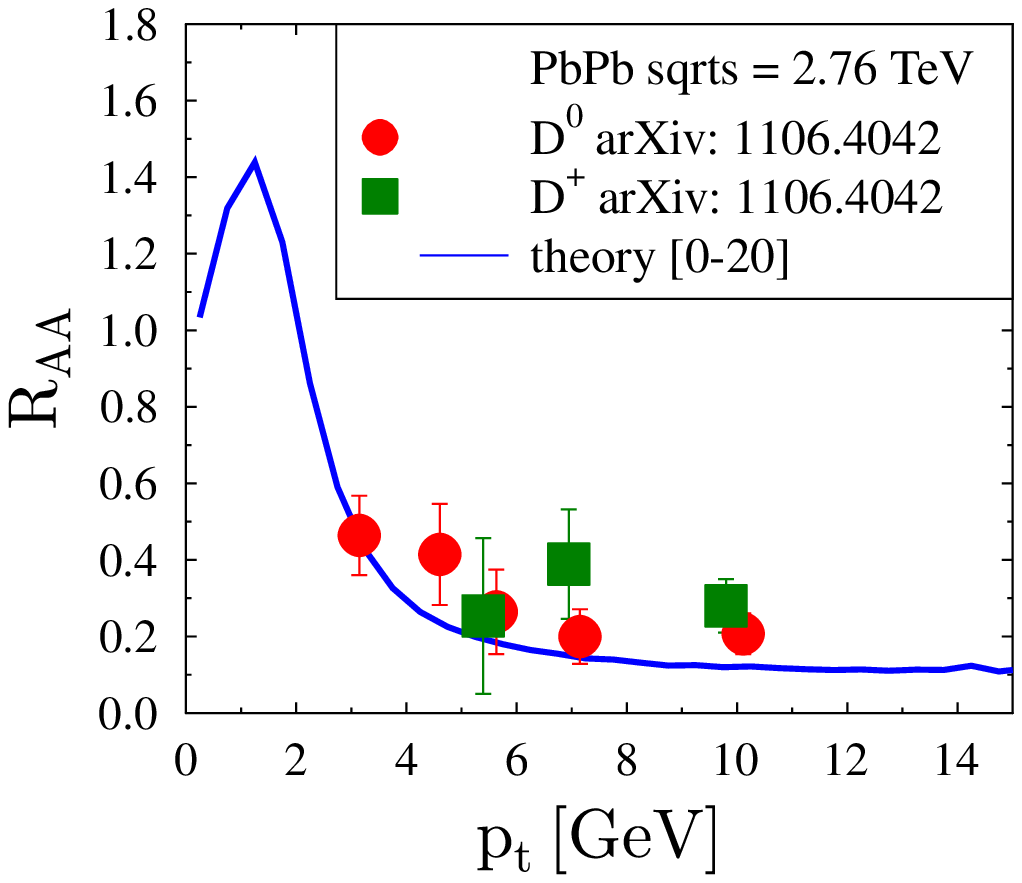,width=0.45\textwidth}
\epsfig{file=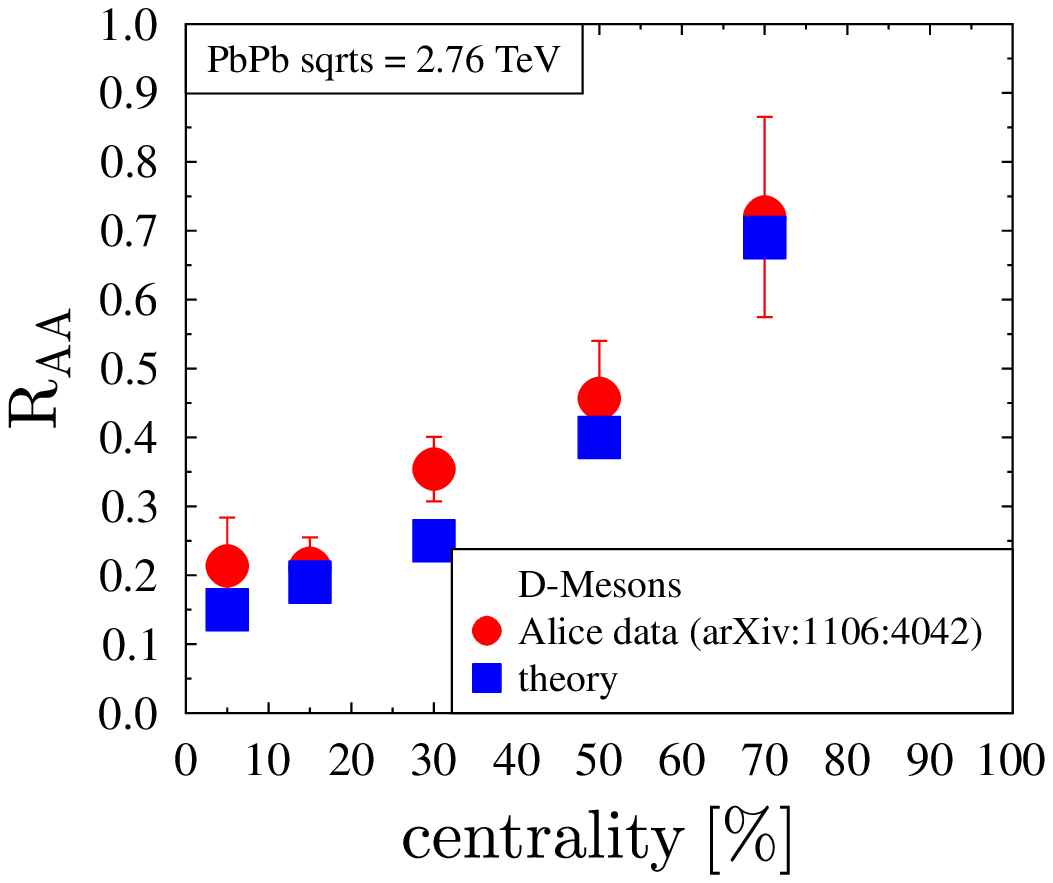,width=0.45\textwidth}
\end{center}
\label{dat1}
\caption{(Color online)Left:  $R_{AA}$ as a function of $p_t$ for 0-20\% centrality, right: centrality dependence of 
$R_{AA}$ . We compare data from the Alice collaboration  \cite{Dainese:2011mw} with our prediction.}
\end{figure}

\begin{figure}
\begin{center}
\epsfig{file=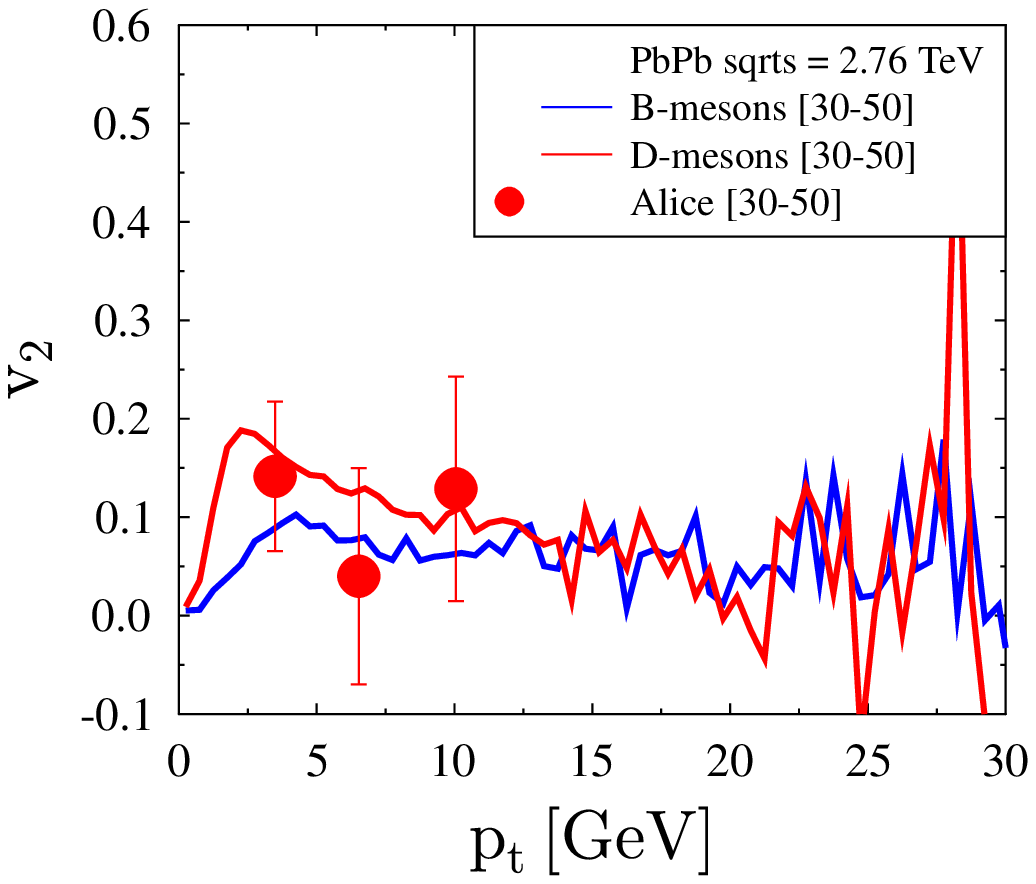,width=0.45\textwidth}
\epsfig{file=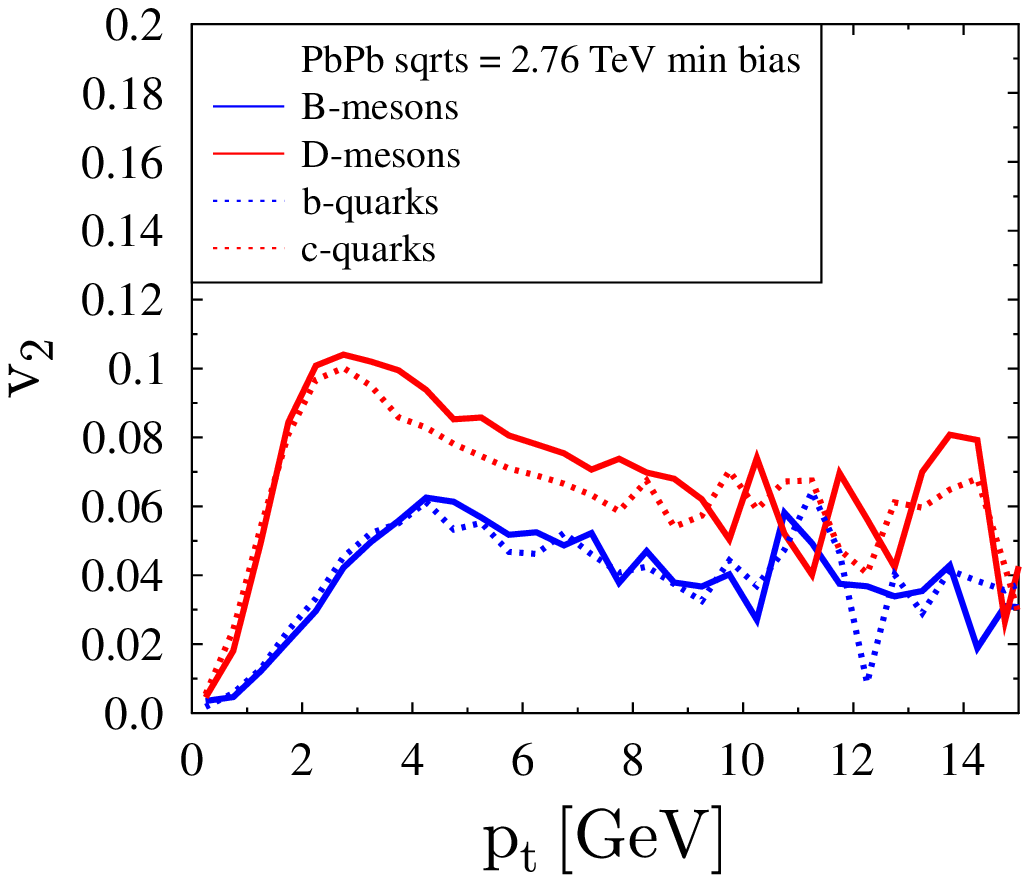,width=0.45\textwidth}
\end{center}
\label{dat2}
\caption{(Color online) $p_t$ dependence of $v_2$. On the left hand side we compare our calculations for D and B meson
for [30-50\%] centrality with the experimental data shown as this conference \cite{}, on the right hand side we display $v_2$
for minimum bias separately for c-quarks and D-mesons and b-quarks and B-mesons, respectively.}
\end{figure}
The scalar product is defined as $p_ap_b= \frac{p_a^+p_b^-+p_a^-p_b^+}{2}-p_{at}p_{bt}$ and $q^2=t \approx q_t^2$. In this coordinate system x is given by $k^+ = x p_a^+$ and represents the relative longitudinal momentum fraction of the gluon  with respect to
the incoming heavy quark. In this coordinate system $|M_{SQCD}|^2$ has a very simple form:
\be
|M_{SQCD}|^2= g^2 D^{QCD} 4(1-x)^2 |M_{elast}|^2\left(\frac{\vec
k_t}{k_t^2+x^2m^2}-\frac{\vec k_t-\vec q_t}{(\vec q_t - \vec
k_t)^2 +x^2m^2}\right)^2
\label{qcdem}
\ee
with the color factor $D^{QCD}=C_A*C_{el}^{qq}=\frac{2}{3}$.
The first term in the bracket describes the emission from the incoming heavy quark line, the second term the emission from
the gluon. This shows that in light cone gauge and in this coordinate system in leading order of $\sqrt{s}$ the matrix element for the emission from the light quark do not contribute.
In the case of massless quarks we recover the squared matrix element of Gunion and Bertsch (GB) of ref.\cite{Gunion:1981qs}.

\section{Results}
Having the matrix elements we can calculate the cross section of the elastic and radiative collisions of the heavy quarks with the plasma particles.  At RHIC we found that the agreement is best when we multiply all cross section with a constant K factor of 0.6. A K factor of one is also compatible with the data but at the limits of the error bars. These cross sections are embedded in the plasma expansion as described in refs. \cite{Gossiaux:2008jv,Gossiaux:2009mk,Gossiaux:2009hr}. 
Fig. \ref{pt} displays the $p_t$ dependence of $R_{AA}$ at midrapidity for different centrality bins and for c and b quarks separately. Charm quarks, being lighter, suffer a larger energy loss than bottom quarks and are therefore pushed
more toward low $p_t$. Below a centrality of 40\% $R_{AA}$  does not change substantially. At small momenta we see an enhancement. There the energy loss accumulates the heavy quarks. 
For large $p_t$, shown in the bottom part of fig. \ref{pt},  radiative collisions act differently than elastic collisions. If we employ only elastic collisions (model E, with a K factor of 2) we see an increase of $R_{AA}$ with $p_T$
whereas for elastic and radiative collisions (with a K-factor of 0.6)  $R_{AA}$ remains almost flat. If we include the Landau Pomeranchuck Migdal effect which suppresses radiation we would expect a moderate increase of $R_{AA}$ with increasing pt. For comparison we display as well the calculation for the RHIC data which matched the experimental results.

Fig. 3 shows the comparison of our calculations with $R_{AA}$ Alice data  \cite{Dainese:2011mw}. On the left hand side we display $R_{AA}$ as a function of $p_t$ of [0-20\%] centrality. The calculations follow closely the experimental data. On the right hand side we see $R_{AA}$ for mesons with $p_t > 6 \ GeV$ as a function for the centrality. Also here we see a good
agreement between theory and experiment

Fig. 4 show the comparison of our calculations with recent $v_2$ Alice data  \cite{sqm}. We see that at low $p_t$ $v_2$ for B-mesons is substantially smaller than for D-mesons.  This is again the consequence of the smaller mass of the c-quarks
which can more easily absorb the $v_2$ of the light plasma particles with whom they collide during the expansion. We
see that the prediction of our model (the data have been presented for the first time at this conference when the calculations have been already performed) agrees with the experimental value in between the error bars. The right hand side highlights
the difference of $v_2$ between b and c quarks at intermediate $p_t$. This difference is inherent in the model and may therefore serve as a verification if perturbative QCD is the right theory to describe the data. Whereas the $v_2$ of D-mesons is slightly higher than that of the c-quarks due to the hadronisation, the heavy B-meson has practically the same $v_2$ as the b-quark before hadronisation. 

In conclusions we have shown that pQCD like models which include a running coupling constant as well as a infrared regulator of the gluon propagator in the  elastic cross section which is based on hard thermal loop calculations reproduce
the LHC data as they reproduced the RHIC data. The model predicts different $v_2$ values for D- and B-mesons as well
as an increase of $R_{AA}$ in central collisions with $p_t$ for $p_t$ larger than 15 GeV. The model can therefore be verified by future experimental data. The results show that collisional as well as radiative energy loss is necessary to describe the data. Both contribute to $R_{AA}$ in a comparable way. In this analysis we used the hydrodynamical model of Heinz
and Kolb. It remains to be seen how other models for the expansion of the plasma change the numerical values of $R_{AA}$
and $v_2$. Studies of different expansion scenarios as well as of the influence of the Landau Pomernschuk Migdal effect are under way.

\end{document}